# The gasdynamic evolution of spherical planetary nebulae

## Radiation-gasdynamics of PNe III

**Garrelt Mellema**[*]

Sterrewacht Leiden, P.O.Box 9513, 2300 RA Leiden, The Netherlands
Internet: mellema@strw.LeidenUniv.nl

**Abstract.** Using a radiation-gasdynamics code the evolution of spherical planetary nebulae is followed, while taking into account the evolution of central star and the fast wind. These models show the importance of ionization fronts for the structure of planetary nebulae, especially for the so called multiple shell nebulae (MSPN). It is shown that the outer shell is formed by the ionization front while the inner shell is swept-up by the fast wind. These models explain the emission profiles of the outer shells as well as their various kinematic properties. Because they are shaped by the ionization front these outer shells only give indirect information on the AGB mass loss history. The models indicate that typical MSPN structures point to mass loss variations during the AGB phase. The ionization also leads to a stalling of the expansion of the nebula, leading to nebulae with expansion ages lower than their evolutionary age. Values for ionized mass and Zanstra temperatures are derived from the models.

**Key words:** Hydrodynamics – Methods: numerical – Stars: post-AGB – Planetary nebulae: general

## 1. Introduction

It is a well established fact that planetary nebulae are formed through a combination of gasdynamic and radiative processes. The gas lost by the star as an AGB red giant (the 'slow wind') feels both the influence of the UV photons from the star and the ram pressure from the fast wind emanating from this future white dwarf. This means that realistic gasdynamic modelling requires the inclusion of a large portion of radiation physics. Because the central star evolves, these processes should be treated in a time, density, and temperature dependent manner. The numerical radiation-gasdynamics method described in Frank & Mellema (1994a) (henceforth Paper I, see also Mellema 1993,

Chap. 3) is capable of doing this. Although some assumptions still have to be made, this method allows gasdynamic modelling in a fashion that was not possible before.

PN image surveys (Balick 1987; Schwarz et al. 1992) show that most PNe are aspherical. Nevertheless this paper is dedicated to spherical models for PNe. The treatment of aspherical models can be found in Mellema (1993) (Chaps. 5 and 6), Frank & Mellema (1994b) (Paper IV) and future papers. There are very good reasons for not discarding the spherical models. The main reasons are that they reveal the basic evolutionary processes and are simple to do. The latter is true in various ways. Because they are one-dimensional models they require less computational effort. Furthermore, they suffer less from numerical problems. They are also simpler to analyze. This last point can be illustrated by the fact that the information from one whole run can be presented in a single grey scale plot. All this means that one dimensional models allow a quick overview of models and the influence of various input parameters.

On the observational side one can classify a rough 20% of PNe as spherical (12 of the 51 in Balick's (1987) sample), and one may even extend this set by including some mildly elliptical nebulae, although caution is required in these cases.

### 1.1. Previous analytical work

A very thorough analytical treatment of the problem of the dynamical evolution of spherical PNe was given by Kahn & Breitschwerdt (1990) and Breitschwerdt & Kahn (1990). The basic picture is the interaction between the PN fast wind and the AGB slow wind. In the first of these articles the authors took into account the effects of an accelerating fast wind, concentrating on the change-over from 'momentum-driven' to 'energy-driven' flow. In the momentum-driven phase, the shocked fast wind easily cools and consequently no hot bubble develops. The fast wind is in direct contact with the slow wind and sweeps up a shell with its ram pressure. When the energy content of the fast wind becomes too high, the cooling can no longer keep up with the shock heating and a bubble of hot shocked gas devel-

_Send offprint requests to:_ G. Mellema
[*]    Present address: Mathematics Department, UMIST, PO Box 88, Manchester, M60 1QD, UK

ops. This bubble separates the fast wind from the slow wind and it is the (thermal) pressure of the hot bubble that sweeps up a shell in the slow wind, the energy-driven phase. Kahn & Breitschwerdt found that the change-over from momentum to energy-driven flow takes place at a fast wind velocity of approximately 150 km s$^{-1}$.

In the second article the authors considered the effects of ionizing radiation on this process. They found that ionization of the slow wind before the change-over to energy-driven flow drives the interface between fast and slow wind towards the star, while an outer shock accelerates into the slow wind. When the flow changes to energy-driven a second outer shock is driven into the slow wind.

### 1.2. Previous numerical work

Because of the relative ease of spherical models there have been several previous attempts at numerical computations. The most recent ones are by Marten & Schönberner (1991) and Frank (1994) (henceforth Paper II). Marten & Schönberner used the method of Schmidt-Voigt & Köppen (1987) and the radiation physics contained in this is quite sophisticated. However, their (Lagrangian) gasdynamic method is not particularly accurate. Furthermore, the inner edge of their computations is associated with the contact discontinuity. Hence they need a description of the position of this contact discontinuity in time. This is a non-trivial problem, as can be seen from the analytical work quoted above. Marten & Schönberner's solution is based on energy-driven flow and discards the complications of the change-over from momentum to energy-driven flow. They considered a slow wind consisting of two different mass loss phases and an evolving star and fast wind. The main conclusion of the article is that their models reproduce the observed nebular thicknesses (the width of the nebular ring relative to the size of the nebula). In Marten et al. (1993) some more observational aspects were treated, such as brightness distribution and expansion velocities, showing that despite the difficulties their method can produce quite useful results.

In Paper II the same treatment for the radiation processes as here was used (i.e. the method described in Paper I), but in combination with a different gasdynamics solver, the FCT/LCD method (Boris & Book 1973,1976; Icke 1991). In it a rather extreme situation is considered: as in Marten & Schönberner the slow wind consists of two phases and the central stars evolves, but its fast wind remains constant with a velocity of 2000 km s$^{-1}$. Consequently the nebula is always energy-driven. In this case the evolution of the PN as a sequence of four phases: pre-ionization, H-ionization, He-ionization, and shock break-out phase. Different phases are associated with different types of PNe: single shell, double shell, and double shell with rapidly expanding outer shells. Although the constant fast wind case allows the study of some interesting basics of PN evolution, it is probably more realistic to assume that the fast wind becomes more energetic during the evolution from a red giant to a hot white dwarf.

Consequently there is room for improvement, especially in including the effects of the change-over from momentum to energy-driven flow. In this paper I combine the best properties of both models mentioned above. Using an accurate gasdynamics solver (the Roe solver, see Mellema et al. 1991) together with the radiation/ionization physics method from Paper I, I look at the interaction of an evolving fast wind with a slow wind. The evolution of the central star is also taken into account.

The paper is organised as follows. In Sect. 2 I describe my treatment of the slow wind, fast wind and central star. Section 3 contains a description of the gasdynamic evolution of the models. Observational aspects (images and kinematics) are treated in Sects. 4 and 5. In Sect. 6 some other observational aspects (Zanstra temperature and ionized mass) are described. Section 7 contains a summary of the conclusions.

## 2. Initial and boundary conditions

The combined radiation-gasdynamic code requires a range of input parameters. These can be divided into three ingredients: the slow wind, the fast wind, and the stellar spectrum.

### 2.1. The slow wind

Numerically, the slow wind constitutes the initial condition. The grid is filled with the assumed slow wind density, velocity, and temperature profile and the fast wind and UV photons from the star interact with it.

Physically, the slow wind is the remnant of the mass loss from the star as an AGB red giant. Typical parameters for mass loss on the AGB are $10^{-7} - 10^{-4}$ M$_\odot$ yr$^{-1}$ for the mass loss rate and $5 - 25$ km s$^{-1}$ for the velocity. The most common velocity is 15 km s$^{-1}$ (see e.g. Blommaert 1992). The velocity of the slow wind is relatively unimportant for the models, since it is much less than the final fast wind velocity. In all models it is taken to be 10 km s$^{-1}$.

Much more crucial is the density distribution in the slow wind. This is determined by the mass loss history on the AGB. A constant mass loss rate at constant velocity results in a $1/r^2$ radial density law. It is almost certain that mass loss on the AGB is not constant, but the actual time dependence is the subject of much debate. Based on statistical arguments Baud & Habing (1983) suggested that the mass loss should dramatically increase in the final phase of the AGB. By linking mass loss rates to certain stellar properties, recent stellar evolution calculations have shown complex time dependencies for the AGB mass loss rates (Vassiliadis & Wood 1993; Blöcker, private communication). In these calculations the mass loss rates increase towards the end of the AGB, but in a pulse-like way, i.e. even towards the end of the AGB there are periods of relatively low mass loss rates. During the whole AGB there are short ($< 1000$ years) mass loss bursts, linked to thermal pulses.

It must be said that there is little or no solid physics behind these mass loss calculations, and consequently they should be used with some caution. On the other hand, they are the only theory available and observational constraints are few.

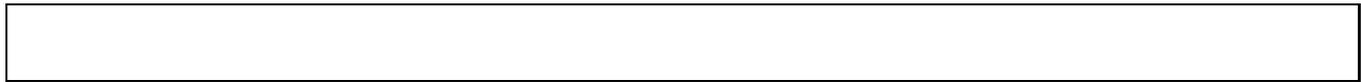

**Fig. 1.** Typical slow wind density distribution (left) and corresponding AGB mass loss history (right). This is the initial density profile of run B, see Table 1

Here I use a two component slow wind, consisting of a superwind phase of high mass loss rate and an AGB wind phase of lower mass loss rate. Both phases are assumed to have the same velocity, $v_{slow}$. The superwind is assumed to have lasted a time $t_{super}$ and consequently extends from the inner edge of the grid ($r_0$) to $r_{super} = r_0 + v_{slow}t_{super}$. The mass loss rate in this period is assumed to be constant and consequently a $1/r^2$ density law applies. Outward from $r_{super}$ the grid is filled with the AGB wind. For this second phase again a $1/r^2$ radial density dependence is used. The two slow wind phases are separated by a transition phase, which is assumed to have lasted a time $t_{trans}$. In this transition zone the two slow wind densities laws are smoothly joined up using (arbitrarily) a cosine function.

To summarize (see also Fig. 1): the superwind extends from $r = r_0$ to $r_{super} = r_0 + v_{slow}t_{super}$, the transition phase extends from $r_{super}$ to $r_{super} + v_{slow}t_{trans}$, and the AGB wind from there to the edge of the grid.

By taking $t_{super}$ very large (say infinite) an effective one-phase slow wind is obtained. Setting $t_{trans} = 0$ produces the type of slow wind used by Marten & Schönberner (1991) and Paper II.

The initial temperature of the slow wind is not an important quantity. After ionization the temperature is fixed by the heating/cooling balance, before ionization the atomic gas has a low temperature and pressure, which is dynamically negligible. The initial temperature is put at a few hundred Kelvin.

### 2.2. The fast wind

From a numerical point of view, the fast wind forms the inner boundary condition. The properties of the fast wind are put in the first cell of the grid and the interaction with the slow wind determines the subsequent evolution of the flow.

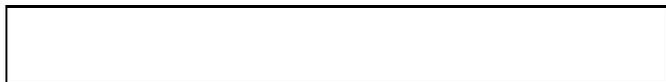

**Fig. 2.** Fast wind mass loss and velocity history. The solid line shows the mass loss rate, the dotted line the velocity. The values are derived from the Kudritzki model for the 0.598 $M_\odot$ Schönberner evolutionary track. For the $k$ parameter of the Kudritzki model the standard value of 0.053 is used

From a physical point of view the fast wind is the mass loss after the AGB phase. At a certain moment (defined as the end of the AGB phase) the high mass loss rate stops. The envelope of the star has been reduced to a fraction of its original mass and starts to contract on the hot stellar core. The typical mass of the remnant envelope is 0.03 $M_\odot$ (Schönberner 1983). The precise conditions for the cessation of severe mass loss are not very well known, nor is the subsequent mass loss behaviour of the post-AGB star. Central stars of PNe are observed to experience mass loss at relatively low rates ($10^{-7} - 10^{-9}$ $M_\odot$ yr$^{-1}$) and high velocities ($10^3 - 4\,10^3$ km s$^{-1}$) (see the review by Perinotto 1993). Stars in between the AGB and PN phases are only recently being studied in a systematic way (see e.g. Oudmaijer et al. 1993; Van der Veen et al. 1993) and apart from the fact that some show circumstantial evidence for mass loss (hot dust), next to nothing is known about their mass loss properties.

Kinematic studies of the CO emission around these stars (so called proto-PN) sometimes reveal fast molecular outflows of up to several hundred kilometers per second (Olofsson 1993). This would indicate that stellar mass loss velocities are at least that high.

The work of Pauldrach et al. (1988) showed that the observed mass loss from hot PN central stars can be adequately explained by the process of radiation pressure on lines. This theory, sometimes known as the CAK theory after the authors who laid the basis for it (Castor, Abbott & Klein 1975), was originally developed for massive O stars. The process that drives the mass loss is the transfer of photon momentum to the gas through absorption by strong resonance lines.

Kudritzki et al. (1989) published an analytical approximation for the (extended) CAK theory. This allows the calculation of the mass loss rate and wind velocity at infinity from the stellar mass, luminosity, and temperature, as well as four parameters (the 'force multipliers') : $k$, $\alpha$, $\delta$, $\beta$. The parameter $k$ is a measure of the number of lines that transfer momentum to the gas. Higher values for $k$ correspond to more lines and result in higher mass loss rates. The value for $\alpha$ corresponds to the saturation of the lines and ranges from 0 to 1; $\delta$ is related to the change of ionization structure throughout the wind; $\beta$ determines the radial velocity variation in the wind. Pauldrach et al. (1988) suggested that the following values should be typical for PNe: $k = 0.053$, $\alpha = 0.709$, $\delta = 0.052$, and $\beta = 1.0$. I use these except that in some cases I change $k$ to experiment with different mass loss rates. Fig. 2 shows the post-AGB wind evolution for a 0.598 $M_\odot$ Schönberner star (see Sect. 2.3) when using the force multiplier values quoted above.

Formally the theory of radiation-driven mass loss is only valid for $T_{eff} > 30\,000$ K. I use it to derive mass loss properties during the whole post-AGB evolution, so for 5000 K and higher. The lack of theoretical and observational understanding of the mass loss process below 30 000 K makes any choice arbitrary. Using radiation-driven mass loss is convenient and results in wind velocities of order of the escape velocity at the surface of the star, not an unreasonable estimate. The mass loss rates

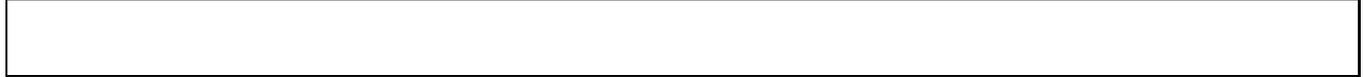

**Fig. 3.** The post-AGB evolution of a star with final mass 0.598 M$_\odot$ according to Schönberner (1983). Left: effective temperature (solid line) and luminosity (dotted line). Right: H (solid line), He$^0$ (dotted line), and He$^{1+}$ (dashed line) ionizing photon rates

derived are a factor 1000 or 10 000 lower than on the AGB and are therefore not in contradiction with the observationally derived drop in mass loss rate after the AGB. I note that a critical comparison between observed and calculated mass loss rates for massive O stars shows that radiation-driven theory can still be off by a factor of ten (Lamers & Leitherer 1993).

### 2.3. The star

When the stellar envelope has been reduced to a small fraction of its pre-red giant mass, the mass loss rate drops precipitously and the star starts to contract. The precise details of this process are unknown and the 'end of the AGB' is an important, but ill-defined concept in stellar evolution calculations.

The subsequent evolution depends critically on the remaining envelope mass[1] (see e.g. Renzini 1989). If a substantial amount remains, the initial evolution is slow, since nuclear fusion will have to reduce the envelope mass up to the point where contraction sets in. If the star has only a small envelope at the end of the AGB, contraction sets in immediately and the evolution is fast. This can be seen in Schönberner's early models (Schönberner 1983). In these he used a temperature criterion to determine the end of the AGB. This leads to relatively high envelope masses for low mass stars. These then take a long time to evolve (see also McCarthy et al. 1990). More recent calculations use a pulsation criterion which leaves them with smaller envelope masses and the post-AGB evolution proceeds much faster (Blöcker & Schönberner 1990). A substantial amount of mass loss after the AGB can also influence the evolutionary time scales.

Note however that it is mainly the time scales that are uncertain, the general character of post-AGB evolution appears to be well understood. Therefore the use of the theoretical evolutionary tracks should produce reasonable models for general PN formation. Only when detailed time scales are considered one should consider the uncertainties in the tracks.

In the runs I present here, I use the evolutionary track leading to a final mass of 0.598 M$_\odot$ taken from Schönberner (1983). Marten & Schönberner (1991) used the same track, but skipped the first 1000 years of evolution, "to get consistency with later calculations ($\ldots$) where larger mass loss rates have been used at the tip of the AGB.". I do not do this, but note that this shows again how uncertain the time scales are.

In Fig. 3 I show the evolution of the effective temperature, luminosity, and the number of H, He$^0$, and He$^{1+}$ ionizing photons produced per second for the 0.598 M$_\odot$ track. These two figures are useful for interpreting the models in this paper.

---
[1] and on whether the star is hydrogen or helium burning when it leaves the AGB (see Wood & Faulkner 1986).

### 3. Gasdynamic evolution

In this and the following sections I describe the results from a selection of runs. The input parameters of these runs are listed in Table 1. In describing the results I will start with the gasdynamics: the evolution of density, velocity, and temperature. The next section shows what these models would look like in the sky and I compare these synthetic images with observations. After that I look at the different kinematic properties of the models: what do long slit spectra and pattern velocities teach us and how does that compare to what is actually observed?

**Table 1.** Parameters for the simulations presented here (see Sect. 2)

| Run | A | B | C | D |
|---|---|---|---|---|
| grid cells | 400 | 400 | 400 | 400 |
| cell size (m) | 2 10$^{13}$ | 2 10$^{13}$ | 2 10$^{13}$ | 2 10$^{13}$ |
| $\dot{M}_{super}$ ( M$_\odot$ yr$^{-1}$) | 10$^{-5}$ | 10$^{-4}$ | 10$^{-4*}$ | 10$^{-4}$ |
| $t_{super}$ (years) | $\infty$ | 1000 | $\infty$ | 1000 |
| $t_{trans}$ (years) | — | 2000 | — | 1000 |
| $\dot{M}_{AGB}$ ( M$_\odot$ yr$^{-1}$) | — | 6 10$^{-6}$ | — | 6 10$^{-6}$ |
| $v_{slow}$ ( km s$^{-1}$) | 10 | 10 | 10 | 10 |
| $T_{slow}$ (K) | 500 | 200 | 200 | 200 |
| stellar mass ( M$_\odot$) | 0.598 | 0.598 | 0.598 | 0.598 |
| $k_{fast}$ | 0.053 | 0.053 | 0.053 | 0.004 |

* at maximum, density falls off as $r^{-3}$

In Fig. 4 I show the development for run A (see Table 1 for an overview of initial conditions). Run A has a one-phase slow wind ($\dot{M} = 10^{-5}$ M$_\odot$ yr$^{-1}$). The fast wind material is derived from the stellar parameters using Kudritzki's algorithm with the standard value for the force multiplier $k$. Figure 4 is a set of time-position diagrams in which grey scales indicate the logarithm of the mass density, radial velocity, and temperature. The white contour in Fig 4a is effectively the boundary between ionized and neutral material. These diagrams give an overview of a whole simulation.

In the early phases one sees the slow wind flowing away from the star with the original 10 km s$^{-1}$. The fast wind material fills the area left by the slow wind, but is not yet powerful enough to sweep up a shell. Cooling is able to bring down the temperature of the shocked fast wind and consequently this is the 'momentum-driven' phase: the fast wind is in direct contact with the slow wind. As the velocity of the fast wind increases, cooling can no longer keep up with it and a hot bubble develops: the flow becomes 'energy-driven'. This development can be most clearly seen in the velocity and temperature diagrams. The change-over happens at $t \simeq 1300$ years at which time the fast wind velocity is $\sim 160$ km s$^{-1}$. This matches exactly the value found in the analytic results of Kahn & Breitschwerdt (1990). No optimization of input parameters was used to obtain these results. Consequently, this is a very nice confirmation of the correspondence between the analytical and numerical work.

**Fig. 4.** Position-time diagrams for run A. Upper left (a): logarithm of the density with the 90% $H^+$ contour. Upper right (b): logarithm of the density with the 90% $He^{+2}$ contour. Lower left (c): logarithm of the velocity. Lower right (d): logarithm of the temperature. The extrema are: $3.8\,10^{-3} < \rho < 3.4\,10^6$ cm$^{-3}$, $3.4 < v < 5.3\,10^3$ km s$^{-1}$, $30 < T < 3.7\,10^8$ K. The very low temperatures occur in pressure undershoots at the inner shock

Around the same time the stellar temperature (see Fig. 3) has increased to such values that an ionization front starts moving out. This front initially moves through the slow wind as a D-front, sweeping up material. Because of the pressure increase in the new ionized slow wind the expansion of the inner shock and contact discontinuity stalls. Breitschwerdt & Kahn (1990) predicted that the contact discontinuity would be pushed back. In this model it merely slows down. The pressure in the hot bubble is directly related to the mass loss rate in the fast wind. For lower mass loss rates I do find that the contact discontinuity and inner shock get pushed back (see run D below).

As the star gets hotter the ionization front turns into an R-front and rapidly ionizes the outer slow wind. The density enhancement produced by the action of the D-front persists and continues to expand after all of the slow wind material has been ionized. It will turn out that this behaviour has important consequences for the appearance of the nebula (see Sect. 4). In the velocity diagram the low velocities behind the D-front are noticeable. This is expected behaviour and is due to the rarefaction wave set up by the D-front (see e.g. Lasker 1966; Marten & Schönberner 1991). The expected temperature increase due to the ionization can be seen in Fig. 4d.

Notice the complex behaviour of the contact discontinuity around the time of the development of the ionization front and the switch over from momentum to energy-driven flow. The development of the hot bubble pushes the inner shock inward. At the same time the expansion of the contact discontinuity slows down, because of the pressure increase in the ionized slow wind.

After some time the contact discontinuity starts moving out again ($t \simeq 3500$ years) and the hot bubble starts sweeping up a high density rim in the ionized slow wind material. This rim gradually becomes more extended. The remnant shell created by the D-front expands at about the same velocity as this rim.

In the position-time diagram for the velocity the acceleration of the fast wind can be clearly seen. As was mentioned above, the rarefaction wave from the D-front causes the low velocities around $t \simeq 2200$ years. Notice also the noisy structure of the velocity in the hot bubble. This was also found in Papers I and II, and indicates a disturbance at the inner shock.

Velocity waves bounce back and forth between inner shock and contact discontinuity. This is possible because of the subsonic character of the flow in the hot bubble and the fact that the one-dimensional simulation provides only one way to go for the waves. In two-dimensional simulations the flow pattern will be much more complex (see Mellema 1993).

The evolution of the temperature shows the relatively low temperatures in the undisturbed fast wind. This is due to its adiabatic expansion (see Frank et al. 1992). The increase in the temperature of the hot bubble can also be seen. This is due to the increase in the fast wind velocity.

Pattern velocities can be derived from Fig. 4 by measuring the angle of gasdynamic waves. A point of reference is given by the fact that angle of 45° with the horizontal axis corresponds here to a pattern velocity of 33 km s$^{-1}$. In discussing the kinematics (Sect. 5) I will come back to these velocities.

Figure 5 shows in the same way as Fig. 4 the results of a simulation for a two-phase slow wind (run B). The main difference here is the blow-out of the superwind into the AGB wind after the D-front has moved through ($t \simeq 3500$ years). Because of the density difference between the two slow wind phases the remnant of the D-front accelerates and the density in the now ionized superwind part quickly drops. At the contact discontinuity not much changes with respect to the one-phase slow wind case. The same slow down can be seen at the moment ionization sets in. Notice the slower expansion of the contact discontinuity than in run A. This is due to the higher density in the slow wind. Because of this the nebula stays ionization bounded for a longer time. Only at $t \simeq 3500$ years ($T_{eff} \simeq 35\,000$ K) the $H^+$ front breaks out. The $He^{2+}$ front breaks out at $t \simeq 6000$ years ($T_{eff} \simeq 85\,000$ K). Notice again in the velocity diagram the rarefaction waves around $t \simeq 2000$ years, and the noise in the hot bubble velocity.

Run C has a one-phase slow wind with the density in the slow wind changing as $r^{-3}$. This corresponds to a more gradual increase of AGB mass loss rate in time ($\propto 1/t$ with $t$ counting backwards from the end of the AGB). As can be seen from the position-time diagrams (Fig. 6), this run basically shows the same type of evolution as the previous two runs: an ionization D-front forms, breaks out, and is followed by a rim swept up by the fast wind. The main difference with runs A and B is a much faster expansion in the later phases. This is because of the much more rapid density fall-off here.

To see what a very weak fast wind does, I show in Fig. 7 the time-position diagrams for a run with the $k$ parameter for the Kudritzki model chosen such that the mass loss rate stays low all the time (run D, see Table 1). Mass loss rates are $10^{-9}$ M$_\odot$ yr$^{-1}$

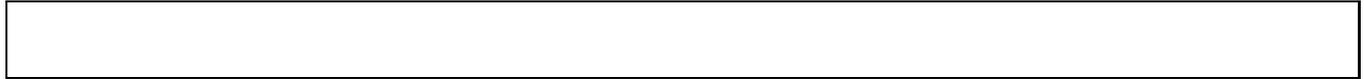

**Fig. 5.** Position-time diagrams for run B. Upper left (a): logarithm of the density with the 90% H$^+$ contour. Upper right (b): logarithm of the density with the 90% He$^{+2}$ contour. Lower left (c): logarithm of the velocity. Lower right (d): logarithm of the temperature. The extrema are: $1.6\,10^{-3} < \rho < 3.4\,10^7$ cm$^{-3}$, $-18 < v < 9.0\,10^4$ km s$^{-1}$, $18 < T < 1.3\,10^9$ K. The very low temperatures occur in pressure undershoots at the inner shock

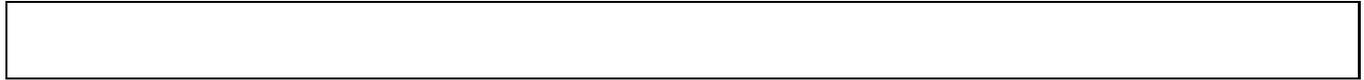

**Fig. 6.** Position-time diagrams for run C. Upper left (a): logarithm of the density with the 90% H$^+$ contour. Upper right (b): logarithm of the density with the 90% He$^{+2}$ contour. Lower left (c): logarithm of the velocity. Lower right (d): logarithm of the temperature. The extrema are: $7.4\,10^{-4} < \rho < 4.0\,10^7$ cm$^{-3}$, $1.9 < v < 4.9\,10^4$ km s$^{-1}$, $10 < T < 3.7\,10^8$ K. The very low temperatures occur in pressure undershoots at the inner shock

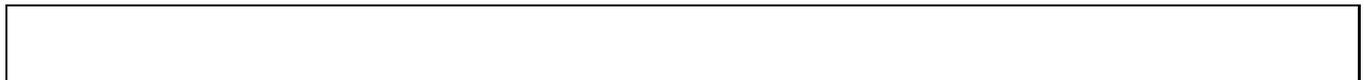

**Fig. 7.** Position-time diagrams for run D. Upper left (a): logarithm of the density with the 90% H$^+$ contour. Upper right (b): logarithm of the density with the 90% He$^{+2}$ contour. Lower left (c): logarithm of the velocity. Lower right (d): logarithm of the temperature. The extrema are: $3.6\,10^{-2} <$ density $< 3.4\,10^7$ cm$^{-3}$, $-16 < v < 5.5\,10^3$ km s$^{-1}$, $10 < T < 1.2\,10^7$ K. The very low temperatures occur in pressure undershoots at the inner shock

in the early phases and drop to about $10^{-10}$ M$_\odot$ yr$^{-1}$ in the later stages. The fast wind velocity increases in the usual fashion. From Fig. 7 it is clear what happens: at the time the ionization sets in the higher pressure of the ionized matter pushes back the contact discontinuity and inner shock, and succeeds in quenching them to the smallest scale the model can handle. In other words: the whole nebula backfills. This was predicted by Breidschwerdt & Kahn (1990). The effect can also be seen in the velocity diagram, where the white area shows material actually flowing back to the star. The result is a nebula without a hole in the middle. Notice that not the whole nebula collapses: the D-front disturbance keeps on travelling outward. The changing density distribution leads to a temporary recombination of some of the gas around 2000 years. Further out, where the gas is less dense, the recombination does not occur. This shows that the model correctly handles time dependent ionization.

The temperature diagram shows how the hot bubble initially forms, but is totally quenched by the pressure from the ionized slow wind.

These four runs illustrate that there are two distinct interactions that determine the shaping of PNe: the ionization (D) front and the fast wind. In the runs shown here, these two are always clearly separated. In runs A to C the fast wind becomes important after the action of the ionization front, in run D it never becomes important, and in the models in Paper II the constant (2000 km s$^{-1}$) fast wind sweeps up a rim before the ionization front starts moving out. Note however that in this last case eventually the same configuration emerges as in runs A to C (compare Fig. 5 from Paper II with my Figs. 4 – 7).

A very general pattern thus emerges, in which the ionization front sets up more extended high density regions and the fast wind the more confined ones (rims). In the next sections I look at what observables this produces.

## 4. Evolution in images

The program allows the construction of synthesized observations. I will present these in two sections. This section deals with the emission line images, the next section with the kinematics. Before showing the images from the simulations, I first discuss the observational side.

### 4.1. Observational aspects of round PNe

Round PNe in Balick's (1987) sample are NGC 1514, 1535, 2392, 3587, 6894, 7139, IC 972, 1454, 3568, A 30, BD30°3639, and He 1-4. Of these, NGC 1514, 3587, and IC 972 are classified as amorphous. NGC 3587 (the Owl nebula) is probably not a spherical PNe, since the two 'eyes' are a typical feature of 'spindle' shaped nebulae seen at an inclination of about 60° (Icke et al. 1992; Frank et al. 1993). NGC 1535, 2392, 6894, IC 3568, He 1-4, and probably IC 1454 have what Balick (1987) calls 'inner haloes', Chu et al. (1987) 'attached shells', and Frank et al. (1990) 'shells'. I will call them (following Paper II) 'envelopes'. The remainder (NGC 7139, A30, and BD30°3639) appear to have sharp edges. NGC 7139 may not be round since the H$\alpha$ and [OIII] images show a more barrel shaped structure in the inner parts. A30 is most likely an old PN, i.e. its central star is on the cooling track. Abundance differences between the inner and outer parts suggest a 'born-again' situation (see Iben et al. 1983; Jacoby & Ford 1983). BD30°3639 is listed by Chu et al. (1987) as having an envelope, but this seems to be an erroneous classification. Balick et al. (1992) report that the very faint emission they found around BD30°3639 is most likely due to scattered light in the telescope.

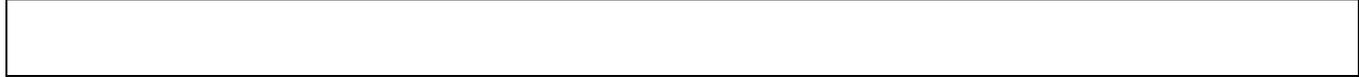

Envelopes are a feature also shown by a large portion of the elliptical PNe. Frank et al. (1990) list other clear cases of envelopes as NGC 2022, 2610, 3242, 6369, 6826, 6857, 7009, 7354, 7662, IC 289, and M 2-2, all early or middle ellipticals in Balick's list.

Chu et al. (1987) did a study of multiple shell PNe (MSPNe). They differentiate between type I (detached) and type II (attached) shells. What I call envelopes correspond to their type II shells. They find that the typical ratio of envelope to rim radii is 2.0 (range 1.17 – 2.8) and of envelope to rim surface brightness 0.25 (range 0.15 – 0.5). Notice that this last ratio is seriously influenced by seeing. They estimate that at least half of all PNe contain multiple shells.

Frank et al. (1990) did a study of these envelopes and found that they can often be described as 'linear', that is the emission profiles often fall off linearly in the radial direction. Figure 8 shows a few examples. Based on the results of some non-radiative gasdynamic simulations, the authors interpreted the linear profiles as evidence for interaction between two slow wind phases: a high mass loss superwind and a lower mass loss red giant wind.

Most of the round PNe listed above show small scale structures, suggesting a clumpy or lumpy structure. The character of the one-dimensional simulations inhibits modelling of these details and I will consequently concentrate on the more global properties: sharp edges, surrounding envelopes, amorphous structure, and differences between images in various lines.

### 4.2. Images from the models

In Fig. 9 I show a collection of cross cuts across the Hα images for run A. One sees initially a compact ionization bounded nebula. At later times this becomes a density bounded double shell nebula.

The compact ionization bounded phase lasts for a short time. PNe in this phase are young, i.e. their stars have low effective temperatures ($T < 30\,000$ K) and the nebula is small. It is very likely that BD30°3639 is in this phase, since Kaler & Jacoby (1991) list its effective temperature as 27 000 K (Stoy method) or 30 000 K (Zanstra method using Hβ) and its diameter is estimated at 0.025 pc (using an optical diameter of 7."5 and a distance of 0.7 kpc, see Acker et al. 1992). For run A the ionization bounded phase ends when the star reaches a temperature of 18 000 K, a very low temperature for a PN. In Kaler & Jacoby's list of effective temperatures for low-excitation PNe, the lowest temperature is 22 000 K. The usual explanation for this is that dust initially blocks the nebular emission. On the other hand, observations indicate that nebulae should be ionization bounded up to stellar temperatures of 40 000 K (Kaler & Jacoby 1991). This points to either a denser slow wind than in case A, or a higher absorption by slow wind material, e.g. by dust, or a lower luminosity star. In a run identical to A, but with a five times denser slow wind, the nebula remains ionization bounded up to a stellar temperature of 30 000 K, as it does in run B. At that time the nebula has a diameter of 0.12 pc. But since the size of a model nebula is closely connected to the uncertain stellar evolutionary time scale and the observationally derived size depends on the distance, it is hard to draw any conclusions based on individual PN size (see McCarthy (1990) for a discussion of sizes and time scales; in the section on kinematics I will come back to some of these issues).

According to the models presented here the formation of a surrounding envelope should happen to almost all PNe. The H ionization front is initially trapped, but eventually breaks out. This break out leads to the formation of the envelope, the bright inner rim being caused by the fast wind. Only when no or a very weak fast wind is present, will the formation of the rim not take place (see below). It is therefore comforting to see that more than half of the round PNe in Balick's sample show a surrounding envelope. In due time the envelope will become much fainter than the rim and will no longer be observable. Marten et al. (1993) also reported the formation of surrounding envelopes in their numerical models.

The envelopes formed in run A have a typical two phase structure: an inner envelope attached to the bright swept up rim and an outer envelope. In the early phases the inner envelope is brighter on the outside than on the inside. In the Frank et al. (1990) terminology this would be a 'crown'. At later stages the inner envelope shows a linear slope before changing over to the outer envelope. The outer envelope shows at all stages a power-law like radial fall off. The brightness ratio of the inner envelope in the early phases is about 50% of the rim dropping to about 20% at later stages. For the outer envelope it stays around 5%. If one compares this to the observed brightness profiles the match is not impressive (compare Figs. 8 and 9). The ratio of the rim and inner envelope radii is about 1.6, a reasonable result. The brightness ratio is somewhat on the large side but not unreasonable. Worse, outer envelopes are typically not observed and the typical linear drop does not show up in the models. Part of the mismatch can be explained by saying that most observed images do not show absolute brightness. If this is true the horizontal axis in Fig. 9 should be shifted upward and a more reasonable correspondence to observations can be obtained. Very deep CCD observations by Balick et al. (1992)

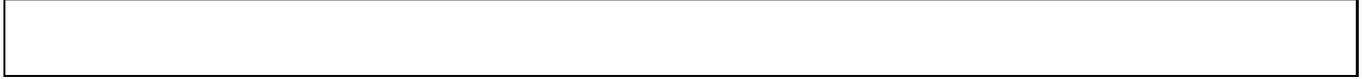

**Fig. 9.** Cross cuts through synthesized Hα images for run A at various times The surface brightness is in units of $10^{-11}$ erg cm$^{-2}$s$^{-1}$arcsec$^{-2}$ for a nebular distance of 1 kpc. Note the shapes of the envelopes

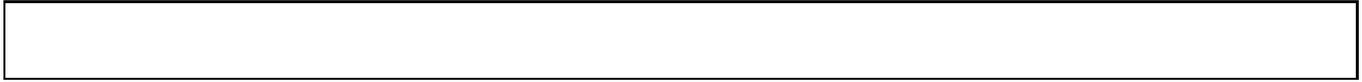

**Fig. 10.** Central crosscuts through the Hα images of run B. The surface brightness is in units of $10^{-11}$ erg cm$^{-2}$s$^{-1}$arcsec$^{-2}$ for a nebular distance of 1 kpc. Note the slopes and edges of the envelopes

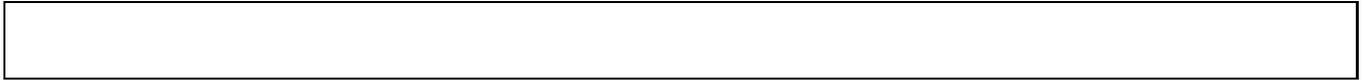

**Fig. 11.** Central crosscuts through the Hα images of run C. The surface brightness is in units of $10^{-11}$ erg cm$^{-2}$s$^{-1}$arcsec$^{-2}$ for a nebular distance of 1 kpc. Note the slopes and edges of the envelopes

show that there sometimes is faint, previously undetected emission around PNe.

One can ask whether there are initial conditions that result in envelopes which correspond better to the observed ones. In particular, does a discontinuous slow wind —as used by Marten & Schönberner (1991) and Paper II— work better?

In Fig. 10 I show some Hα surface brightness profiles for run B, which has a two-phase slow wind (see Table 1). The envelopes are of right relative size and surface brightness with respect to the rim. They fit the observations better in the sense that they lack the extended outer envelopes. The surface brightness drops considerably beyond the edge of the envelope. Linear drops can be discerned beyond $t = 7129$ years, and typical 'shoulder' patterns (as seen for instance in NGC 3242, 6826, 7354) at $t = 4278$ years.

Taking an $r^{-3}$ radial density distribution for the slow wind (corresponding to a smoother increase of the AGB mass loss in time), also produces nice linear envelopes (see the run C brightness profiles in Fig. 11). These envelopes also reproduce the observed sizes and brightness ratios and appear to reproduce the observed linear profiles somewhat better.

This shows that the lack of very extended envelopes very likely indicates that the density drops off faster than $r^{-2}$ in the slow wind. But whether the density profile contains discontinuities or is very gradual cannot be easily deduced from the brightness profiles. These results also show that ionization effects are essential for understanding the properties of envelopes.

All this means that one should be careful in using the surrounding envelope to derive information about the mass loss on the AGB. I find that the action of the ionization front modifies the properties of the slow wind considerably. Notice for instance that in run B the original superwind region has an extent of 0.032 pc, but at $t = 4000$ years it has become as large as 0.1 pc. Hence it is dangerous to directly use the surrounding envelopes of PNe to derive properties of mass loss on the AGB.

All figures above show the Hα emission that one would observe. Because some other key emission lines, such as the [OIII] and [NII] forbidden lines are also calculated by the code, it is also possible to compare different emission line images. As an example I show in Fig. 12 images in Hα, [OIII], and [NII] for run B at $t = 3327$ years. The central star is at a temperature of 33 000 K. One sees how the [NII] traces the ionization front, whereas the [OIII] mainly shows the bright inner rim. The Hα picture shows some of both. This set looks very much like He 1-4, also shown in Fig. 12. Unfortunately its central star is too faint to allow a temperature determination, so there is no good way to check whether the evolutionary stage of He 1-4 indeed corresponds to this model.

As in Paper II I find that the HeII images stay compact up to the point that the star reaches a temperature of 75 000 K. At that point a substantial part of the outer parts get ionized and the HeII pictures begin to look like the Hα pictures.

In run D the mass loss rate for the fast wind was kept very low. In the previous section it was shown that this led to a collapse of the central cavity after ionization sets in. In Fig. 13 I show two Hα images from this run. At $t = 1901$ years the image shows a limb brightened nebula, because the ionization front is still trapped in the superwind part of the slow wind. The next image shows the phase after that ($t = 2852$ years). The Hα picture is now very centrally condensed. This suggests that low fast wind mass loss rates may be used to explain some of the amorphous PNe. Unfortunately, the classification 'amorphous' turns out to be rather arbitrary. NCC 3587 was classified by Balick (1987) as amorphous, but in fact turns out to be an excellent example of a 'spindle'-shaped nebula seen at an angle (see Icke et al. 1992; Frank et al. 1993). Of the two remaining round amorphous PNe in Balick's sample, NGC 1514 may also be a misclassification. It certainly has hints of an interesting morphology. Even IC 972, which is by far the most amorphous round PN in Balick's sample, may have a cavity in the middle of the Hα picture. The only other candidates in Balick's sample

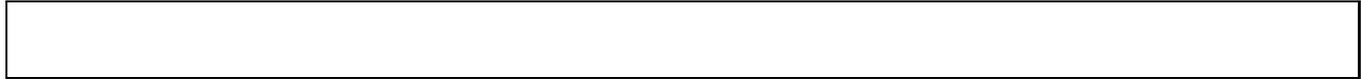

**Fig. 12.** North-south crosscuts across Hα, [OIII], and [NII] images of He 1-4 (data from Balick 1987) and the same for run B at $t = 3327$ years. In both cases the surface brightness in all three lines has been normalized to the maximum in the Hα picture

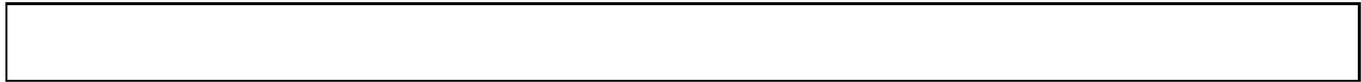

**Fig. 13.** Synthesized Hα images for run D at two times

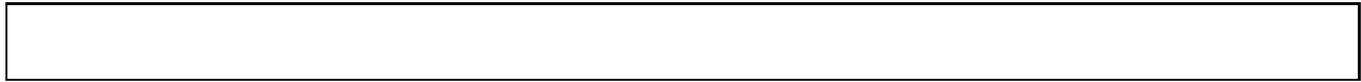

**Fig. 14.** Long slit echelle data from Chu and Jacoby (private communication) for IC 3568 and NGC 2022 in [OIII]. Notice that IC 3568 has the surrounding envelope expanding faster than the bright rim and that NGC 2022 has the rim expanding faster than or as fast as the envelope

are IC 4593 and NGC 6210, which definitely look centrally condensed in all lines.

Hence, one can conclude that centrally condensed nebulae are at best very rare. This means that very low fast wind mass fluxes (lower than a few times $10^{-10}$ M$_\odot$ yr$^{-1}$) are very uncommon or at least were preceded by much higher values. These very low values are sometimes quoted for fast wind mass loss rates (Perinotto 1989). More recent results show mass loss rates ranging from $1.0\,10^{-6}$ to $1.4\,10^{-9}$ M$_\odot$ yr$^{-1}$ (Perinotto 1993).

## 5. Evolution in kinematics

Using the velocity and emission information from the models I can construct kinematic data from the models for comparison with the observations. Kinematic data for PNe typically come in two types, depending on how much of the spatial information is used: long slit echelle spectra with one-dimensional spatial information (a position-velocity diagram), and imaging Fabry-Perot spectra (e.g. the Taurus system) with two-dimensional spatial information. The type that is still most popular is the long slit spectrum. I will use this representation.

### 5.1. Observed kinematics

The basic kinematic property of PNe is their expansion. For most nebulae the slit spectra are roughly elliptical, indicating expansion (see for instance Fig. 14 and Pottasch 1984, p. 128). The distortion from perfect ellipticity must be due to non-spherical motions, and hence need to be explained by the multi-dimensional models (see Paper IV). The models in this paper can be used to study expansion velocities and the kinematic differences between rims and envelopes.

Expansion velocities of the round PNe from Balick's sample range from 7.8 to 53 km s$^{-1}$ (data taken from the Acker et al. (1992) catalogue). This is a large range and consequently it is not to be expected that very much can be derived from this. Furthermore, one should be careful with these non-uniformly obtained data. Acker et al. list an expansion velocity of 7.8 km s$^{-1}$ for IC 3568, whereas data from Chu (1989, and private communication) show that velocities up to 20 km s$^{-1}$ occur. This is mainly because these data probe the fainter outer parts.

There has been some discussion whether expansion velocities increase with kinematic age. The correlation is not very good, but ages are highly uncertain since they are derived from (distance dependent) sizes and assumptions about the expansion of the nebula (Pottasch 1984, p. 131). A somewhat better correlation is found for expansion velocities and central star temperatures (Heap 1993), which for the early phases of PN evolution indicates an age-expansion velocity relation (since central star temperature initially increases with age).

The most important kinematic property on which the models should shed light is the observed fact that in some PNe with surrounding envelopes, the envelopes are expanding faster than the rims (Chu 1989, see also Fig. 14 for an example). This led Chu to the conclusion that the interacting winds model in some cases cannot explain PNe, since during interaction the outer envelope should be moving slower than the expanding rim. The envelopes in Chu's sample are found to be expanding with velocities between 35 and 50 km s$^{-1}$. This points to a dynamical origin for the envelopes, since undisturbed slow wind material is expected to be expanding with velocities of $5-25$ km s$^{-1}$.

Paper II showed one way in which faster expanding envelopes can be generated within the interacting winds model: in the case of a two-phase slow wind a blow out of the forward shock into the lower density material can generate this situation. However, even for the very powerful fast wind of Paper II, this blow out only occurs in the later phases of PN evolution. The three PNe with envelopes expanding faster than rims mentioned by Chu (1989) (IC 3568, NGC 6826, and NGC 6891) are all thought to be low temperature (52 000, 33 000, and 35 000 K respectively) and hence young PNe (see Kaler & Jacoby 1991). The one other example I know of (NGC 3242) fits this picture

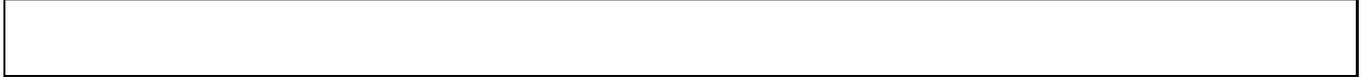

**Fig. 15.** Synthesized [OIII] velocity-position diagrams for run A at various times. The velocities have been smoothed with a Gaussian of $4 \, \mathrm{km \, s^{-1}}$ FWHM

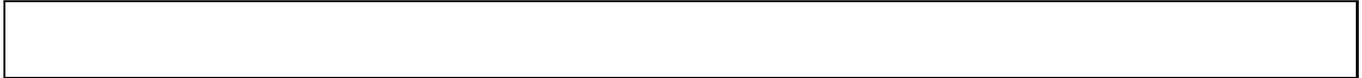

**Fig. 16.** Hα emissivity (solid line) and velocity (dotted line) from run A against radius for two times. At $t = 2852$ years the bright rim is expanding slower than the outer parts of the surrounding envelope; at $t = 4753$ years, it is expanding faster

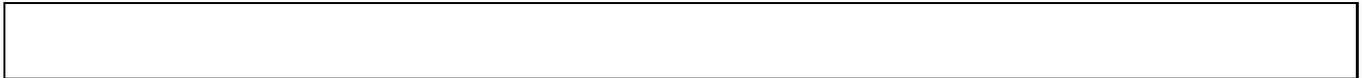

**Fig. 17.** Synthesized [OIII] velocity-position diagrams for run B at various times. The velocities have been smoothed with a Gaussian of $4 \, \mathrm{km \, s^{-1}}$ FWHM

less well since its effective temperature is 89 000 K. Nevertheless there seems to be a strong tendency for fast envelopes to form around young PNe, something that is not seen in the models from Paper II.

## 5.2. Kinematics from the models

In Fig. 15 I show synthesized [OIII] long slit spectra for run A at six different times. In the first frame the nebula is still ionization bounded and expanding at about 20 km s$^{-1}$. In the second frame a surrounding envelope has formed (cf. Fig. 9). The envelope is expanding at 20 km s$^{-1}$, the inner rim is expanding at 10 km s$^{-1}$, so here we have the outer envelope expanding faster than the inner rim, exactly the situation found by Chu (1989) for IC 3568, NGC 6826, and NGC 6891. To understand how this can happen I plot in Fig. 16 the Hα emission and the radial velocity. The figure shows that because of the action of the ionization front the outer parts of the envelope have acquired a large velocity. The inner parts are moving slower than the inner rim and consequently there is no conflict with the interacting winds picture (see also Fig. 4c).

At later times the inner rim accelerates and starts comoving with the envelope (Fig. 15c,d) and ends up moving faster ($v_{\mathrm{exp}} = 32$ km s$^{-1}$) than the envelope ($v_{\mathrm{exp}} = 20$ km s$^{-1}$). So, all of Chu's cases occur within one simulation. Note that run A is the simplest case: the slow wind corresponds to constant mass loss on the AGB. In these simulations the phase in which the outer envelope expands faster than the inner rim occurs early on, typically around the time that the hydrogen ionization front breaks out. As stated above, this is consistent with the evolutionary state of the three nebulae Chu mentions for this case (IC 3568, NGC 6826, and NGC 6891). Note however that the velocity of the envelope in the synthesized data (20 km s$^{-1}$) is lower than in the observations (35 – 50 km s$^{-1}$).

To offer more insight in the kinematic behaviour of various models I present in Fig. 17 the synthetic long slit spectra for run B. Because of the break-out of the envelope into the low density AGB wind, the envelope has a higher velocity (38 km s$^{-1}$) than in run A. This is more in line with observations, confirming that models with AGB mass loss variations better reproduce the observed characteristics of PN envelopes. Run C produces similar values for the envelope expansion velocities.

Because of the larger velocity and densities in the envelope, the velocity of the rim stays lower than that of the surrounding envelope for a much longer time. Only at $t = 8555$ years ($T_{\mathrm{eff}} = 142\,000$ K), does the rim expansion velocity reach the envelope velocity.

Because of the different way the surrounding envelope forms, its kinematic properties will always be distinct from that of the inner rim. Therefore it should in principle be possible to use observations to determine which initial conditions are most realistic. For instance, looking at Chu & Jacoby's small sample one finds that faster expanding envelopes occur preferentially among young PNe. This would allow one to rule out models in which the velocity of the rim stays lower than that of the envelope for a substantial part of the evolution (as in run B). Unfortunately, the current sample of PNe with envelopes and good kinematics is too small to warrant such a conclusion.

## 5.3. Expansion velocities and ages

Observed kinematic properties of PNe can be used to derive secondary quantities. One of these is the so called expansion age, in which the PN radius divided by the observed kinematical expansion velocity gives a PN age.

From the figures in the Sect. 3 it is clear that the pattern expansion of the nebulae does not proceed uniformly. The results in Sect. 5.2 show that the kinematic velocities also vary a lot. How does this influence the expansion ages?

In Fig. 4 an angle of 45° corresponds to a pattern expansion of 33 km s$^{-1}$. It is clear that the nebula expands on average slower than this. Using the kinematic velocities from Fig. 15 and the sizes of the brightest part of the nebula from Fig. 9, one can check that for the six times in Fig. 15 the expansion age/real age combinations are 1600/1901, 3000/2852, 2400/3802, 3100/4753, 3300/5703, and 3900/6654 years. Note that picking the radius of the bright rim for $t \geq 3802$ years temporarily brings down the expansion age.

These values show that initially the correspondence between expansion and evolutionary age is pretty good, but at later times the expansion age drops considerably below the evolutionary age (3900 against 6654 years). When taking the outer edge of the envelope as the size of the nebula, the difference is less extreme (4900 against 6654 years), but is still there.

I find this to be true in all my models. Because of the stagnation of the expansion of the contact discontinuity around the time of ionization of the slow wind, the expansion ages are systematically lower than the evolutionary ages, especially in the later stages. McCarthy et al. (1990) find exactly the reverse effect when they compare dynamical (expansion) age and evolutionary age: most PN central stars have too large nebulae around them. They suggest two possible solutions for this. The first one, rapid ionization of the slow wind, is not supported by my models. Their second solution is that the problem lies with the evolutionary time scales. As was pointed out in Sect. 2.3, it is likely that there are problems with the detailed time scales in the theoretical stellar evolution calculations. Note however also that McCarthy et al. use spectroscopic distances which on average are larger than distances based on other methods (see the review by Terzian 1993). More distant, i.e. larger nebulae would of course worsen the discrepancy.

The numerical model presented by Marten et al. (1993) shows higher pattern expansion velocities than dynamic velocities (leading to higher expansion ages than real ages). The correspondence between their model and the ones presented here is generally quite good, except for this result. It may be that the precise definition of the velocities and the position at which they are measured are crucial here. This is evidently a problem that deserves attention from both theoretical and observational sides.

## 6. Other observational aspects

In the previous two sections I concentrated on emission line images and kinematic data. This section deals with more indirect observables: the Zanstra temperatures and ionized masses.

### 6.1. Zanstra temperatures

Assuming the nebula to be optically thick allows one to estimate the temperature of the central star. By comparing the emission in hydrogen recombination lines (as a measure of the total number of ionizations) with the stellar emission at an almost line-free wavelength, the stellar temperature can be derived (Zanstra 1931; Harman & Seaton 1966; Pottasch 1984, p. 166). The same method can also be used with $He^{+2}$ recombination lines. Often there is a difference between the two temperatures derived from H and He lines, the so called 'Zanstra discrepancy'. This is commonly attributed to either optical depth effects (the nebulae are optically thick to $He^+$ ionizing radiation, but not to H ionizing radiation), or deviations of the stellar spectrum from a blackbody spectrum. Which of these explanations is to be preferred is a long standing debate (cf. Pottasch 1992, who argues for non-blackbody spectra and optically thick nebulae, and Kaler & Jacoby 1989, who prefer the optical depth explanation and find no evidence for significant deviations from black body spectra).

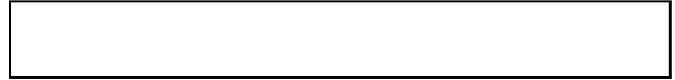

**Fig. 18.** Effective temperature and H and He Zanstra temperatures for run B

All the models shown in this paper become optically thin to H and eventually even to $He^+$ ionizing radiation. This of course influences the Zanstra temperatures derived from them, as can be seen in Fig. 18 where I plot the stellar effective temperature and the two Zanstra temperatures against time for run B. As long as the nebula is optically thick, the Zanstra temperatures agree fairly well with the effective temperature of the star. After the time that the ionization front breaks out, the corresponding Zanstra temperature stays roughly the same and hence falls below the effective temperature of the star. It is remarkable that after the ionization break-out the derived temperature does not change very much. This would indicate that the derived Zanstra temperature is either an estimate of the true effective temperature or of the effective temperature at the time that the nebula became optically thin.

The character of my models (and those of Marten & Schönberner 1991 and Frank 1994) naturally places them at the optical depth side of the Zanstra argument. Since no dust is contained in the models, the actual optical depths may be higher. A longer lasting superwind phase can also keep the nebula optically thick for a longer time. Because of this, it is hard to really use these models to distinguish between the two explanations, but in view of the good correspondence between the models and observed images and kinematic data, my tentative conclusion is that the Zanstra discrepancy can be attributed to optical depth effects.

### 6.2. Ionized masses

Another indirect observable often used is the ionized mass. As is derived by Pottasch (1984) (his Eq. V-7), the ionized mass in solar masses can be estimated from the $H\beta$ flux using

$$M_{\text{ion}} = 11.06 \, F(\text{H}\beta) \, d^2 \, (T/10^4)^{0.88} \, n_e^{-1} \quad M_\odot \,, \tag{1}$$

with $F(\text{H}\beta)$ the $H\beta$ flux in units of $10^{-11}$ erg cm$^{-2}$ s$^{-1}$, $d$ the distance in kiloparsec, and $n_e$ the electron density in cm$^{-3}$. The density and temperature are derived from forbidden line ratios, mainly from [OII] and [SII]. An alternative way to derive $M_{ion}$ is from 6 cm fluxes and angular sizes (see Zijlstra 1990; Pottasch 1992).

From the models, values for $M_{ion}$ can be derived using the same method and be compared to the real ionized mass on the grid. A complication arises in the choice for the values of $T$ and $n_e$. My models do not allow a direct forbidden line analysis. That is why I use a [OII] emission weighed average for both the density and temperature. It is not immediately clear that a full forbidden line analysis would yield these values. It is mainly the deviation in the electron density that contributes to this uncertainty, since the temperature in the nebular regime varies too little to be really important in Eq. (1).

In Fig. 19 I show the ionized mass derived for run B from the H$\beta$ analysis, as well as the real ionized mass on the grid. It is clear that while the nebula is optically thick for H ionizing radiation, the derived ionized mass is a good indication of the real ionized mass. Because of the expansion of the ionization front, the ionized mass increases in time. After the nebula becomes optically thin, the real ionized mass on the grid slowly becomes less. This is because ionized matter is now flowing off the grid. In reality, it would stay constant. In this phase the mass obtained from the H$\beta$ analysis is found to underestimate the mass by a factor 2. Since my estimate for density does not correspond directly with an observational one, this quantitative result cannot be carried over directly to the observations. It does make clear that the choice for the density is important, and severe under- and overestimates for the mass are possible.

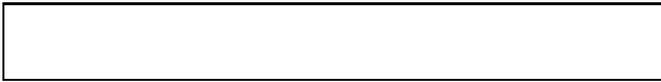

**Fig. 19.** Ionized masses against time for run B. The solid line is the ionized mass contained in the grid, the dotted line the number derived from a synthesized H$\beta$ flux determination

Marten & Schönberner (1991) also calculated the ionized mass in their models. They find a continuous increase of ionized mass in time, even after the break-out of the ionization front. The reason for this is that they only consider the mass swept-up by the shock when calculating the ionized mass and discard the ionized undisturbed slow wind material. The motivation for this is that the surface brightness of the undisturbed slow wind material is too low to be measured. This is true in the cases when the H$\beta$ flux is measured using surface brightness studies, but not when global absolute H$\beta$ fluxes are used. Since modern studies (see e.g. Cahn et al. 1992) use these more reliable global fluxes, it would seem justified to take into account the ionized material in the undisturbed slow wind, as I did in calculating the numbers in Fig. 19.

From observations there are both claims that the ionized mass keeps increasing with radius (Pottasch 1992), as well as that the relation levels off for larger radii (Meatheringham et al. 1988; Dopita & Meatheringham 1990). This controversy is related to the one about the Zanstra discrepancy, since it is again the optical depth which determines whether the ionized mass keeps increasing or levels off in time. For the $M - R$ relation the observational uncertainties are large. Which radii and/or densities should be used? Are the fluxes absolute? Furthermore, the nebulae may be clumpy with neutral material getting slowly ionized in dense clumps.

My models agree with the nebulae becoming optically thin, and hence a levelling off of the $M - R$ relation, but the same caveats as for the Zanstra temperatures apply here.

## 7. Conclusions

Although spherical models for PN formation have been constructed before, it is striking how much work there still can be done using these models. The key to this a radiation gasdynamic code that is capable of faithfully calculating flows which contain extremely steep jumps in density, velocity, and pressure under realistic heating and cooling conditions.

In a way one could see work like this as the follow-up to all the years of photoionization modelling. The radiation physics contained in the program used to construct the models in this paper falls short of anything a full-grown photoionization code can do, but because of the gasdynamics which is included, it can shed light on many issues related to PN evolution.

The conclusions reached can be summarized as:

1. The numerical models confirm the analytical results of Kahn & Breidschwerdt (1990) and Breidschwerdt & Kahn (1990). In particular:
   a. The change-over from momentum-driven to energy-driven flow occurs at a fast wind velocity of about $150 \, \mathrm{km \, s^{-1}}$.
   b. The ionization of the slow wind slows down the expansion of the contact discontinuity or for weak fast winds actually drives it back towards the star.
   c. The ionization of the slow wind leads to the formation of a first outer shock. The change-over from momentum-driven to energy-driven results in the formation of a second shock.

2. The first outer shock can be identified with the edge of the surrounding envelope in PNe with attached envelopes. The models suggest that unless the evolution of the fast wind proceeds radically different from the Pauldrach et al. (1988) models, all PNe should at some stage have such a surrounding envelope. Marten et al. (1993) also found the formation of a surrounding envelope in their models.

3. The second outer shock can be identified with the bright inner rim.

4. Surrounding envelopes with sharp edges can be formed in many ways. They do not *necessarily* indicate time dependence of the slow wind (i.e. AGB mass loss variations). However, the correspondence with observations (both in brightness profiles and expansion velocities) is better for a time dependent slow wind. The distinction between a sudden and more gradual mass loss variations is hard to make.

5. Because the surrounding envelope and the rim are formed in different ways they can have different kinematic properties.

Especially, the (outer parts of the) envelope may expand faster than the rim. Since the envelope expansion velocity stays roughly constant and the rim accelerates, faster expanding envelopes are mainly expected in young nebulae. This is confirmed by observations (Chu 1989).

6. Because the expansion of the rim stalls during the time of ionization of the slow wind, the expansion (dynamical) age of nebula in the models falls behind the (real) evolutionary age. That is, nebulae are generally too small for their expansion age. Marten et al. (1993) found exactly the opposite result in their numerical models. Why this is, is hard to say. It may be due to differences in the precise definition of the various velocities and sizes that play a role.

7. Too low mass loss rates or velocities in the fast wind lead to the collapse of the nebula when ionization sets in. If fast wind velocities can be estimated by the escape velocity at the stellar surface, fast wind mass loss rates lower than $10^{-9}$ $M_\odot$ yr$^{-1}$ at the moment of ionization would be ruled out, except maybe for centrally condensed nebulae such as IC 4593 and NGC 6210.

8. The models favour explaining the Zanstra discrepancy by optical depth effects, and a levelling off of the ionized mass-radius relation. Note however that this conclusion is dependent on the initial conditions used.

Because of the wide range of plausible parameters, only a fraction of possible models has been treated here. Consequently I expect that further work will give new insight in the problems of PN formation. Particular areas of interest are the time scales, the surrounding structures (envelopes and outer haloes), as well as the question whether a considerable fraction of PNe is optically thick to H ionizing radiation.